\documentstyle[pra,aps,graphicx,epsfig]{revtex}

\newcommand{\ds}{\displaystyle}
\newcommand{\beq}{\begin{equation}}
\newcommand{\eeq}{\end{equation}}
\newcommand{\bit}{\begin{itemize}}
\newcommand{\eit}{\end{itemize}}
\newcommand{\barr}{\begin{array}}
\newcommand{\earr}{\end{array}}
\newcommand{\rr}{{\bf r}}

\newcommand{\F}{{\cal{F}}}
\newcommand{\G}{{\cal{G}}}
\newcommand{\N}{{\cal{N}}}

\mathchardef\ogon="012C%
\newcommand{\as}{a\kern-0.22em\lower.40ex\hbox{$_{\ogon}$}}
\newcommand{\As}{A\kern-0.22em\lower.40ex\hbox{$_{\ogon}$}}
\newcommand{\es}{e\kern-0.22em\lower.40ex\hbox{$_{\ogon}$}}
\newcommand{\Es}{E\kern-0.22em\lower.40ex\hbox{$_{\ogon}$}}
\newcommand{\ee}{{\mbox{e}}}
\newcommand{\ii}{{\mbox{\footnotesize{i}}}}
\newcommand{\II}{{\mbox{i}}}
\author{Radka Bach,$^{1}$ Miros{\l}aw Brewczyk,$^2$ Kazimierz Rz{\as}{\.z}ewski$^1$}
\address{
   $^1$Center for Theoretical Physics and College of Science,\\
   Polish Academy of Sciences, Aleja Lotnik\'ow 32/46, 02-668 Warsaw, Poland\\
   $^2$Bia{\l}ystok University, ul. Lipowa 41,
   15-424 Bia{\l}ystok, Poland}
\title{Finite temperature oscillations of a Bose-Einstein condensate in a
two-gas model}

\begin{document}
\maketitle

\begin{abstract}
The temperature dependence of the frequencies of a Bose-Einstein
condensate obtained in experiment has not been fully understood
theoretically. In this paper we present a simplified version of a
two-gas model. A numerically-found ground state of the system is
used for the small-oscillations analysis. In the case of spherical
symmetry a full spectrum of frequencies is found for low orbital
quantum numbers. Avoided crossings that appear in the spectrum
might be the reason for experimentally observed frequency shifts.
\end{abstract}

\section{Introduction}
One of the first experiments done after achieving Bose-Einstein
condensates was measuring its eigenfrequencies
\cite{Cornell:1,Ketterle:1}. These experiments were performed at
very low temperatures, where no thermal atoms were detectable. The
results agreed very well with the predictions based on the
mean-field theory \cite{Clark,You,Stringari,Schlapnikow}. This
success was not repeated with later experiments, which measured
the frequencies of the condensate in full temperature range
\cite{Cornell:2,Ketterle:2}. To explain the experimental results
several models were proposed
\cite{Popow,Hutchison,Dodd:1,Dodd,Stoof,Griffin:1}. To the best of
our knowledge none of these models managed to reproduce
experimentally observed frequency shifts.

In this paper we investigate a simplified version of a two-gas model.
The thermal cloud influence on frequency spectrum is taken into account,
as well as dynamical interaction between the gases. Thermal atoms are treated
in hydrodynamic limit but no mechanism for exchanging the particles
between the phases is provided.
A ground state of such a system is found numerically, and it shows, that
there are qualitative differences with a ground state considered in
other models. For small oscillations linear response theory is applicable
-- it leads to an eigenproblem. In the case of spherically
symmetric trap the resulting equations are solved
numerically.

The paper is organized as follows. In Section 2 the model is furthermore
specified. Numerical strategies are described in Section 3, while Section 4
is devoted to the results.

\section{Two--gas model of BEC at finite temperature}

We develop a simple model of Bose-Einstein condensates at finite
temperatures. In this picture the trapped Bose gas at temperature
below the Bose-Einstein condensation temperature is considered as
a system consisting of two distinct components. One of these
components, the condensed atoms, is described by the generalized
Gross-Pitaevskii equation. The second component, the cloud of
thermal atoms, behaves like a classical ideal gas. We assume that
the relaxation processes within the thermal gas are extremely
fast. Hence, the thermal gas is all the time in the thermal
equilibrium and its temperature can be determined experimentally
by means of its free expansion. The model works in the so called
hydrodynamic regime as opposed to the approach presented in
\cite{Stoof} investigating collisionless regime of parameters.
Similar model was considered in \cite{Griffin:1}.

The number of atoms in each component is preserved, i.e., no
mechanism allowing the exchange of atoms between two factions is
included in this version of the model. 
The fluctuations of the number of atoms in the condensate have
been calculated to scale with the number of atoms as 
$\sqrt{{\cal{N}}_0}$ \cite{Rzaz}.
Therefore their influence is negligible (less than 1\% for
      temperatures below 0.9 $T_c$ in our simulations) since we are 
dealing with ${\cal{N}}$=50 000 atoms.
The relative number of condensed and thermal atoms is taken from the
thermodynamic expression for an ideal gas in the trap. As we know
\cite{fse:1,fse:2}, residual interactions and finite size effects
modify the temperature depletion curve $1-\left( T/T_c\right)^{3}$
in a qualitatively insignificant way.

The model is time--dependent. It is defined by the following set of
equations:
\beq
\begin{array}{@{}rcl@{}}
 \II\hbar\frac{\partial }{\partial t} \Psi_c (\rr,t)& = &
\ds \left[ - \frac{\hbar^{2}}{2 m} \nabla^{\:2} + V_{ext} (\rr,t)
+\; U_0 \left( {|\Psi_c (\rr,t)|}^{\;2} +
2\; \rho_{th}(\rr,t) \right) \right] \Psi_c (\rr,t) \\
 \frac{\partial}{\partial t}  \rho_{th} (\rr,t)& = & \ds - \nabla \cdot
\left[ \rho_{th} (\rr,t)\; {\bf v}_{th} (\rr,t) \right] \\
 m \frac{\partial}{\partial t}  {\bf v}_{th} (\rr,t)& = &
\ds - \nabla \left[
k T \ln{\rho_{th} (\rr,t)} + V_{ext} (\rr,t) +
\frac{m}{2}\; {\bf v}^{\;2}_{th} (\rr,t) +
2 U_0 \;|\Psi_{c}(\rr,t)|^2 \right]  \;
\end{array}
\label{2gas:dep}
\eeq
where $\Psi_c (\rr,t)$, $\rho_{th} (\rr,t)$, and
${\bf{v}}_{th}(\rr,t)$ are the condensate wave function, the
density and velocity fields of the thermal cloud, respectively.

Contrary to the models which consider the thermal part of
the Bose gas as a static cloud \cite{Dodd,2gas:1,2gas:2},
in our approach both components dynamically affect each other.
The influence of the thermal part on the condensed one is
described by a kind of Gross-Pitaevskii equation with an
extra term (the last term on the right hand side of the first
equation of (\ref{2gas:dep})). This term can be considered as a
time--dependent generalization of the mean--field interaction
with the non-condensed density derived in
Hartree-Fock-Bogoliubov-Popov approximation \cite{Griffin:2}.
As has been shown in \cite{Dodd}, going beyond Popov
approximation by taking anomalous averages into account
does not reproduce experimental results. Hence, in this paper we concentrate
on different mechanisms that can lead to frequency shifts.
The second and the third equations of (\ref{2gas:dep}) describe the
thermal cloud as a classical gas confined in the effective
potential formed by the external potential and the condensate
density (the last term in the third equation of (\ref{2gas:dep})).

In the case of no coupling between two components we have
independent oscillations of pure condensate (in accordance
with the Gross-Pitaevskii equation) as well as the thermal
cloud (in a hydrodynamic limit). Assuming no macroscopic
motion of the thermal component, the third equation leads
to the Maxwell-Boltzmann distribution for a gas confined
in the external potential.

Before going into further analysis of the equation (\ref{2gas:dep}), the
trap potential should be specified as an anisotropic harmonic oscillator:
\beq
V_{ext} = \frac{m}{2} \left( \omega_x^2 x^2 + \omega_y^2 y^2 +
\omega_z^2 z^2 \right) =: \frac{m}{2} \sum_i \omega_{x_i}^2 x_i^2
\label{anizotr}
\eeq
Then the external potential defines both the length scale (the size of the
ground state wavefunction) $\alpha=\sqrt{\frac{\hbar}{2m \omega_0}}$ and
the time scale $\tau = \omega_0^{-1}$, where $\omega_0$ is the geometrical
average of trapping frequencies $\omega_0 = \sqrt[3]{\omega_x \omega_y
\omega_z}$. Then the equations (\ref{2gas:dep}) can be expressed in the
following form:
\beq
\barr{@{}rcl@{}}
\II\frac{\partial }{\partial t} \Psi_c (\rr,t)& = &
\ds \left[ - \nabla^{\:2} + 
\sum_{i} \frac{\omega_{i}^{2} x_i^2}{4}
+\; \frac{\gamma}{2} \left( {|\Psi_c (\rr,t)|}^{\;2} +
2\; \rho_{th}(\rr,t) \right) \right] \Psi_c (\rr,t) \\
\frac{\partial}{\partial  t}  \rho_{th} (\rr,t)& = & \ds - \nabla \cdot
\left[ \rho_{th} (\rr,t)\; {\bf v}_{th} (\rr,t) \right] \\
\frac{\partial}{\partial t}  {\bf v}_{th} (\rr,t)& = & \ds
- \nabla \left[
\Omega \ln{\rho_{th} (\rr,t)} + 
\sum_{i} \frac{\omega_{i}^{2} x_i^2}{2}
+ \frac{1}{2}\; {\bf v}^{2}_{th} (\rr,t) +
2 \gamma \;\rho_{c}(\rr,t) \right]  \; \\
\earr
\label{2gas}
\eeq
Now $\omega_{i} \equiv \omega_{x_i}/\omega_0$, $\Omega$ denotes the relative
temperature: $\Omega = 2 kT/\hbar \omega_0$, and $\gamma$ measures the strength
of interactions: $\gamma = 16 \pi a$ ($a$ is dimensionless scattering length).


Stationary solutions of (\ref{2gas}) are found by using the
imaginary--time propagation technique. First, the second and third
equations of (\ref{2gas}) are rewritten in terms of a nonlinear
Schr\"odinger equation, which is allowed provided the thermal
velocity field is irrotational (this assumption was actually
utilized in writing the third equation of (\ref{2gas:dep})). An extra
term appears on the right
hand side of the second equation of
(\ref{NSE}); it is up to the
sign the so called "quantum pressure"
term, known from the hydrodynamic formulation of quantum mechanics
\cite{Madelung}.
\beq
\barr{rcl}
\ds \II \frac{\partial}{\partial t}  \Psi_c (\rr,t)& = &
\ds \left( -  \nabla^{\:2} + 
\sum_{i} \frac{\omega_{i}^{2} x_i^2}{4}
+ \frac{\gamma}{2}\left( {|\Psi_c (\rr,t)|}^{2} +
2 {|\Psi_{th}(\rr,t)|}^{\;2}   \right) \right)
\Psi_c (\rr,t)  \\
\ds \II \frac{\partial}{\partial t}  \Psi_{th} (\rr,t)& = &
\ds \left( - \nabla^{\:2} + 
\sum_{i} \frac{\omega_{i}^{2} x_i^2}{4}
+ \Omega \ln{|\Psi_{th}(\rr,t)|}^{2} +
\frac{\nabla^{2}|\Psi_{th} (\rr,t)|}
{|\Psi_{th}(\rr,t)|} + 
    2 \gamma {| \Psi_c (\rr,t) |}^{2} \right)
\Psi_{th} (\rr,t)
\earr
\label{NSE}
\eeq
To find the stationary solutions of (\ref{NSE}) one has to make
transition to the imaginary time: $t \rightarrow i\;t$.
Assuming, the ground state wave function of a two--gas system
described by (\ref{NSE}) is real, we obtain the set of equations
for real functions $\Phi_{c}(\rr,t)$ and $\Phi_{th}(\rr,t)$
which in the limit $t \rightarrow \infty$ converge to the ground
state wave functions: $\psi_{c}^0(\rr)$ and $\psi_{th}^{0}(\rr)$.
\beq
\barr{rcl}
\ds - \frac{\partial}{\partial \tau}  \Phi_c (\rr,t)& = &
\ds \left( - \nabla^{\:2} + 
\sum_{i} \frac{\omega_{i}^{2} x_i^2}{4}
+\; \gamma \left( {\Phi_c^2 (\rr,t)} +
2\; {\Phi_{th}^2(\rr,t)}   \right) \right)
\Phi_c (\rr,t)  \\
\ds - \frac{\partial}{\partial \tau}  \Phi_{th} (\rr,t)& = &
\ds \left( - \nabla^{\:2} + 
\sum_{i} \frac{\omega_{i}^{2} x_i^2}{4}
+ \Omega \ln{\Phi_{th}^2(\rr,t)} +
 \frac{\nabla^{2}\Phi_{th} (\rr,t)}
{\Phi_{th}(\rr,t)} + 
    2 \gamma {\Phi_c^2 (\rr,t)} \right)
\Phi_{th} (\rr,t)
\earr
\label{imag}
\eeq

\begin{figure} \centering
\begin{tabular}{@{}cc@{}}
\epsfxsize=7.8cm
\epsffile{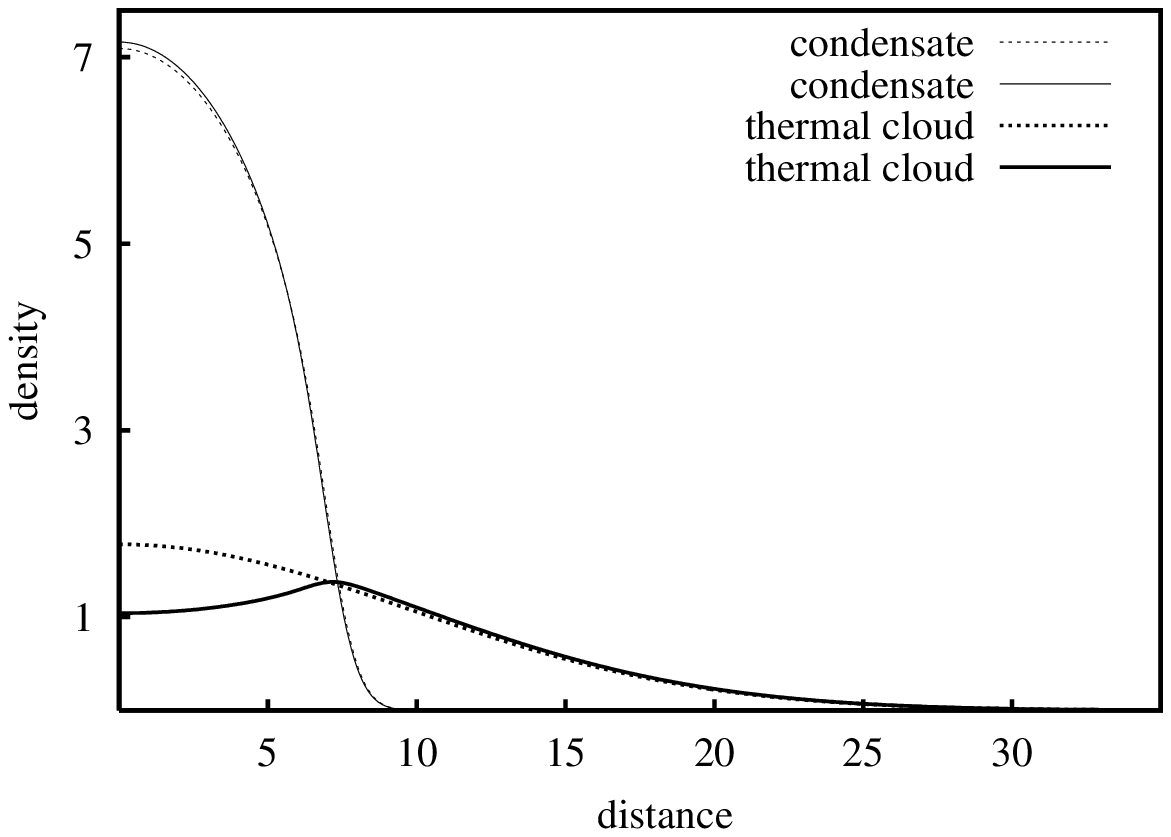} &
\epsfxsize=7.8cm
\epsffile{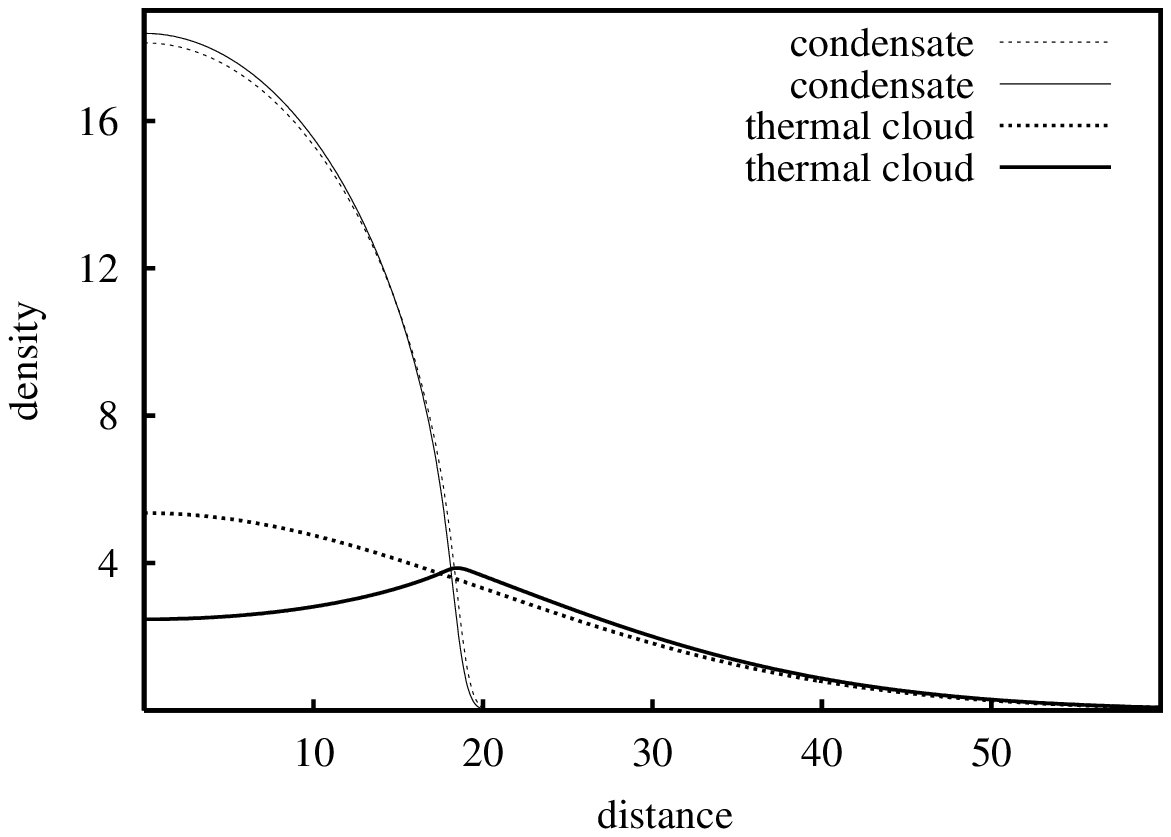} \\
\end{tabular}
\caption{Densities of the ground state in the case of spherical trapping
potential. Solutions of (\ref{2gas}) are
plotted with continuous lines, dotted lines denote the ground state of the
system in the case of no interaction between phases. On the left figure there
is $5 \:000$ atoms, on the right $5\:000\:000$. On both the temperature is
$0.67 \: T_c$.}
\label{rys21}
\end{figure}
Examples of the ground state densities are plotted in figure \ref{rys21}.
If there was no interaction between the phases, we would get a parabola for
the condensate and a Gaussian for the thermal cloud. Taking this interaction
into account, however, results in slight change of the condensates density,
and in great change of the thermal cloud, which is no longer described by
the Gaussian, but by a function with a maximum at the edge of the condensate.

The imaginary--time propagation technique was proved to work in
the case of linear Schr\"odinger equation. We have checked that
it also works for the highly nonlinear problem just defined.
The set of partial differential equations (\ref{imag}) constitutes
an initial--value problem of the parabolic type. We have solved
it by using various numerical methods, depending on the dimensionality
of the system (for a one--dimensional Bose gas the split--operator
technique was used whereas in a three--dimensional case we
applied the Runge-Kutta algorithm).


To find the spectrum of frequencies in the two-gas model, we
use the method of small oscillations around the numerically found ground
state of the system. The assumption of small oscillations can be written:
\beq
\barr{rcl}
\rho_c({\bf r},t) &=& \rho_c^0({\bf r}) + \ee^{- \ii \omega t} \; \delta \!
\rho_c({\bf r}) \label{r:as1} \\
\rho_{th}({\bf r},t) &=& \rho_{th}^0 ({\bf r}) + \ee^{- \ii \omega t} \;
\delta\!\rho_{th}({\bf r})
\earr
\label{r:ansatz}
\eeq
where the common frequency for both phases is already introduced. Of course,
$\rho_c^0(\rr) \equiv \left(\psi_c^0(\rr)\right)^2 $ and
$\rho_{th}^0(\rr) \equiv \left(\psi_{th}^0(\rr)\right)^2 $ denote densities of
the ground state of the condensate and the thermal cloud, respectively.
Substitution these equations into the last two of (\ref{2gas}), linearization
in $\delta \rho_c(\rr)$ and $\delta \rho_{th}(\rr)$, and
elimination of velocity field yields the following:
\beq
\omega^2 \delta \! \rho_{th}({\bf r}) +  \nabla \cdot \left[ {\rho_{th}^0
({\bf r}) \nabla \left( { \Omega \frac{\delta \! \rho_{th}({\bf r})}
  {\rho_{th}^0 ({\bf r})} +
2 {\gamma} \delta \! \rho_c({\bf r}) } \right) } \right] =0
\label{r:chm}
\eeq
It is worth noticing that the equation (\ref{r:chm}) does not allow for an
arbitrary phase shift between oscillations of the condensate and the thermal
cloud. This can be easily seen by writing the equations (\ref{r:ansatz})
in real (not complex) form and following the linearization procedure again.
The resulting equation will have a time-dependent factor in it; setting
this factor constant will lead to allowed phase shift: $\phi=0$ or $\phi=\pi$.
This argumentation is valid locally, which means that for high frequency
modes in some spatial region both gases might oscillate in phase, while in
others out of phase.

As far as condensate is concerned, the small oscillation analysis leads to
the Bogoliubov equations, modified by the presence of the thermal cloud.
First, the second of the (\ref{r:ansatz}) together with the following ansatz:
\beq
\Psi({\bf r},t) = \ee^{-\ii \mu t/\hbar} \left(\sqrt{{\cal{N}}_0}
\psi_c^0({\bf r}) +
u({\bf r}) \ee^{-\ii \omega t} + v^*({\bf r}) \ee^{\ii \omega t}\right)
\label{ansatz}
\eeq
are inserted into the first of the equations (\ref{2gas}). Linearization
procedure and comparing the coefficient standing next to $\ee^{\pm \ii
\omega t}$ results in:
\beq
\barr{rcl}
- {\cal{L}} u({\bf r}) + \frac{\gamma}{2} \rho_c^0 ({\bf r}) v({\bf r})
+ \gamma \sqrt{\rho_c^0} \delta \! \rho_{th}({\bf r}) &=& \omega u({\bf r}) \\
- \frac{\gamma}{2} \rho_c^0 ({\bf r}) u({\bf r}) + {\cal{L}} v({\bf r})
- \gamma \sqrt{\rho_c^0} \delta \! \rho_{th}({\bf r}) &=& \omega v({\bf r})
\earr
\label{r:Bogol}
\eeq
where ${\cal{L}} := -\nabla^2 +\sum_i \frac{\omega_i^2 x_i^2}{4} +
\gamma\left(\rho_c^0({\bf r})+ \rho_{th}^0({\bf r})\right) - \mu$.

It is easily seen, though, that the role of
$\delta \! \rho (\rr)$ in equation (\ref{r:ansatz})
plays $\sqrt{\rho_c^0(\rr)} \: (u(\rr)+v(\rr))$. The Bogoliubov equations
can be transformed so to introduce
$\overline{\delta \! \rho_c} := u+v$:
\begin{equation}
\left({\cal{L}}-\frac{\gamma}{2} \rho_c^0({\bf r})\right) \left\{
\left({\cal{L}} +\frac{\gamma}{2} \rho_c^0({\bf r})\right)
\overline{\delta \! \rho_c} ({\bf r}) - 2 \gamma \sqrt{\rho_c^0 ({\bf r})} \:
\delta \! \rho_{th} ({\bf r})\right\} =
 \omega^2 \: \overline{\delta \! \rho_c}({\bf r})
\label{r:kon}
\end{equation}

The equations (\ref{r:chm}) and (\ref{r:kon}) constitute an
eigenvalue problem:
\begin{equation}
\left( \begin{array}{cc}
    \left({\cal{L}}-\frac{\gamma}{2} \rho_c^0\right)
       \left({\cal{L}} +\frac{\gamma}{2} \rho_c^0\right)&
    - 2 \gamma \left({\cal{L}}-\frac{\gamma}{2} \rho_c^0\right)
      \sqrt{\rho_c^0} \\
      - 2 \gamma \nabla \cdot (\rho_{th}^0 \nabla \sqrt{\rho_c^0}) &
      - \Omega \: \nabla \cdot (\rho_{th}^0 \nabla \frac{1}{\rho_{th}^0})
    \end{array} \right) \left(
    \begin{array}{@{}c@{}}
    \overline{\delta \! \rho_c} \\
    \delta \! \rho_{th}
    \end{array} \right) = \omega^2 \left(
    \begin{array}{@{}c@{}}
    \overline{\delta \! \rho_c} \\
    \delta \! \rho_{th}
    \end{array} \right)
\label{r:problem}
\end{equation}
where ${\cal{L}} := -\nabla^2 +\sum_i\frac{\omega_i^2 x_i^2}{4} +
\gamma\left(\rho_c^0({\bf r})+ \rho_{th}^0({\bf r})\right) - \mu$.


Up to now the considerations were fully anisotropic. For further analysis
however, only the simplest case, the spherical symmetry is assumed.
Then the external
trap is defined: $V_{ext}=\frac{m}{2} \omega_0^2 r^2$. In systems with
spherical symmetry the angular dependence takes form of the spherical
harmonics; it is then possible to seek the global solution in the form:
\beq
\barr{rcl}
\overline{\delta \! \rho_c}({\bf r}) &=& \F(r)
    Y_{lm}(\theta, \phi) \\
\delta \! \rho_{th}({\bf r}) &=& \G(r)
    Y_{lm}(\theta, \phi)
\earr
\eeq
Then the equation (\ref{r:problem}) transforms to:
\begin{equation}
\left( \begin{array}{cc}
    \left(\hat{\cal{L}}_r+\frac{\gamma}{2} \rho_c^0\right)
       \left(\hat{\cal{L}}_r -\frac{\gamma}{2} \rho_c^0\right)&
    - 2 \gamma \:\left(\hat{\cal{L}}_r+\frac{\gamma}{2} \rho_c^0\right)
       \sqrt{\rho_c^0} \\
      - 2 \gamma \left( \frac{1}{r^2} \frac{\partial}{\partial r}
    \left( r^2 \rho_{th}^0 \frac{\partial}{\partial r}\right)
    - \frac{l(l+1)}{r^2} \rho_{th}^0 \right) \sqrt{\rho_c^0} &
    - \Omega \left( \frac{1}{r^2} \frac{\partial}{\partial r}
    \left( r^2 \rho_{th}^0 \frac{\partial}{\partial r} \frac{1}{\rho_{th}^0}
    \right) - \frac{l(l+1)}{r^2} \right)
    \end{array} \right) \left(
    \begin{array}{@{}c@{}}
    \F \\ \G
    \end{array} \right) = \omega^2 \left(
    \begin{array}{@{}c@{}}
    \F \\ \G
    \end{array} \right)
\label{r:problem1}
\end{equation}
where ${\hat{\cal{L}}}_r \equiv - \frac{1}{r^2} \frac{\partial}{\partial r}
\left( r^2 \frac{\partial}{\partial r} \right) + \frac{l(l+1)}{r^2} +
\frac{\omega_0^2 r^2}{4} + \gamma (\rho_c^0+\rho_{th}^0) - \mu$.

\section{Numerical approach}
The problem is first discretized on a lattice. Then the equations
(\ref{r:problem1}) are solved directly by diagonalization of the
resulting matrix. The full spectrum of eigenfrequencies and
eigenmodes is found.

Boundary conditions are set both at the boundary of the lattice and
at the origin.
For large distances we demand that the functions $\F(r)$, $\F'(r)$ and
$\G(r)$ vanish. In the origin the boundary conditions depend on the value of
orbital quantum number $l$.

To obtain these we used the analytical expression
for the condensates wavefunction valid in the Thomas-Fermi limit and in
$T=0$ regime \cite{Schlapnikow}: \label{falTF}
\beq
\F(r) =\sqrt{4n+2l+3} \: \sqrt{\rho_c^0} \: r^l P_n^{(l+1/2,0)}(1-2 r^2)
\eeq
where $P_n^{(l+1/2,0)}$ are Jacobi polynomials \cite{Abramowicz}.
For $l=0$ the boundary conditions are: $\F'(r=0)=0$ and $\F'''(r=0)=0$;
for $l=1$: $\F(r=0)=0$ and $\F''(r=0)=0$.
Of course, they are identical as for the radial equation of the ordinary linear Schr{\"o}dinger equation.

For the thermal cloud, the equation (\ref{r:chm}) in the lack of interaction
between the gases can be solved analytically. The ground state is just
a Gaussian:
\beq
\rho_{th}^0(r) = \frac{{\cal{N}}_{th}}{(2 \pi \Omega)^{3/2}} \exp \left(
- \frac{r^2}{2 \Omega} \right)
\eeq
and the second of the equations (\ref{r:problem1}) takes the form:
\begin{equation}
\Omega \; \G''(r) + \G'(r) \left( \frac{2 \Omega}{r} +r \right) +
\G(r) \left( \omega^2 + 3 - \frac{l(l+1)}{r^2}\Omega \right) = 0
\end{equation}
Two independent solutions are:
\beq
\barr{rcl}
\G^{(1)}(r) &=& \ee^{-r^2/2 \Omega} \; r^l \;
\phantom{}_1 \! F_{1} \left(
\frac{3+l}{2} + \frac{\omega^2}{2}, \frac{3}{2}+l, -\frac{r^2}{2 \Omega} \right) \\
\G^{(2)}(r) &=&  \ee^{- r^2/2 \Omega} \; \frac{1}{r^{1+l}} \;
\phantom{}_1 \! F_{1} \left( \frac{2-l}{2}
+\frac{\omega^2}{2}, \frac{1}{2}-l, -\frac{r^2}{2 \Omega} \right)
\earr
\eeq
where $\phantom{}_1\!F_1$ denotes Kummer's hypergeometric function
\cite{Abramowicz}. The solution $\G^{(2)}(r)$ is singular at $r=0$. For
the solution $\G^{(1)}(r)$ to vanish at $r \rightarrow \infty$ the
frequency must be:
\beq
  \omega = \sqrt{2 n+l} 
\label{3widchm}
\eeq
From the explicit form of the solution the boundary conditions can be
deduced: $G'(r=0)=0$ for $l=0$ and $G(r=0)=0$ for $l>0$, again as in
quantum mechanics.

\section{Results}
All of the calculations were performed for the rubidium condensate
with parameters: $m=1.44 \cdot 10^{-22}$g, $a=5.821$nm, $\omega_0=
2 \pi \cdot 200$Hz and $\N=50\;000$ atoms. This does not mean however,
that the results are specific to this set of parameters only. In our
model there are only two dimensionless parameters describing the system:
$\gamma$ and $\Omega$. The type of atoms comes into play only when $\gamma
\sim a/\sqrt{m}$ is concerned. For the sodium and rubidium condensates
the difference in $\gamma$ is around 8\%.

\begin{figure} \centering
\epsfxsize=14cm
\epsffile{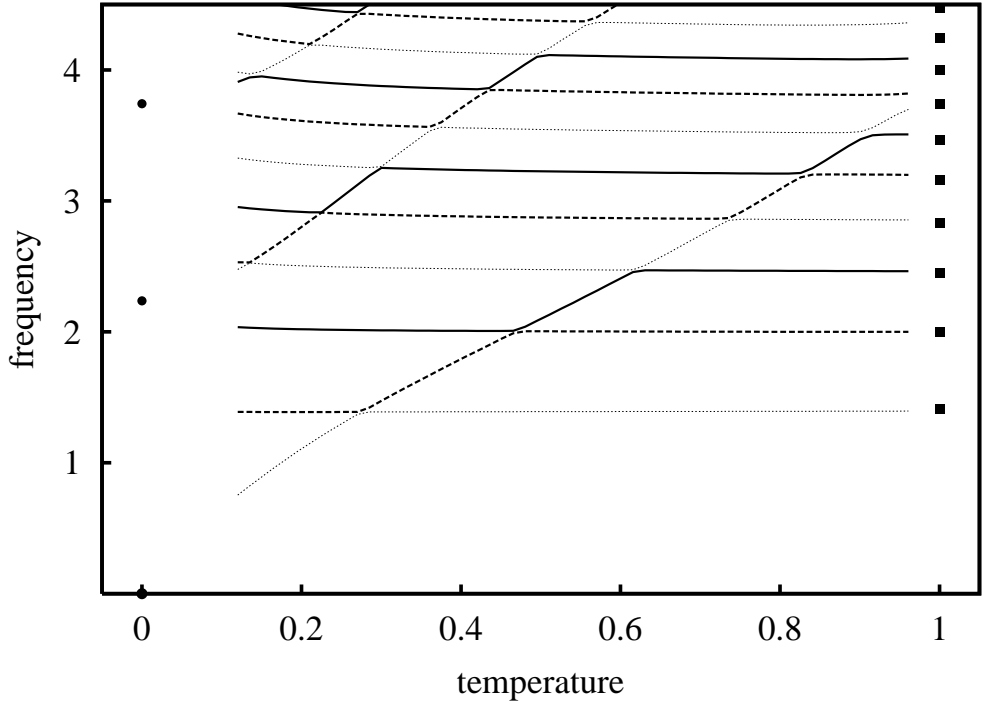}
\caption{Eigenfrequencies versus temperature for $l=0$.}
\label{rys41}
\end{figure}

\begin{figure} \centering
\epsfxsize=14cm
\epsffile{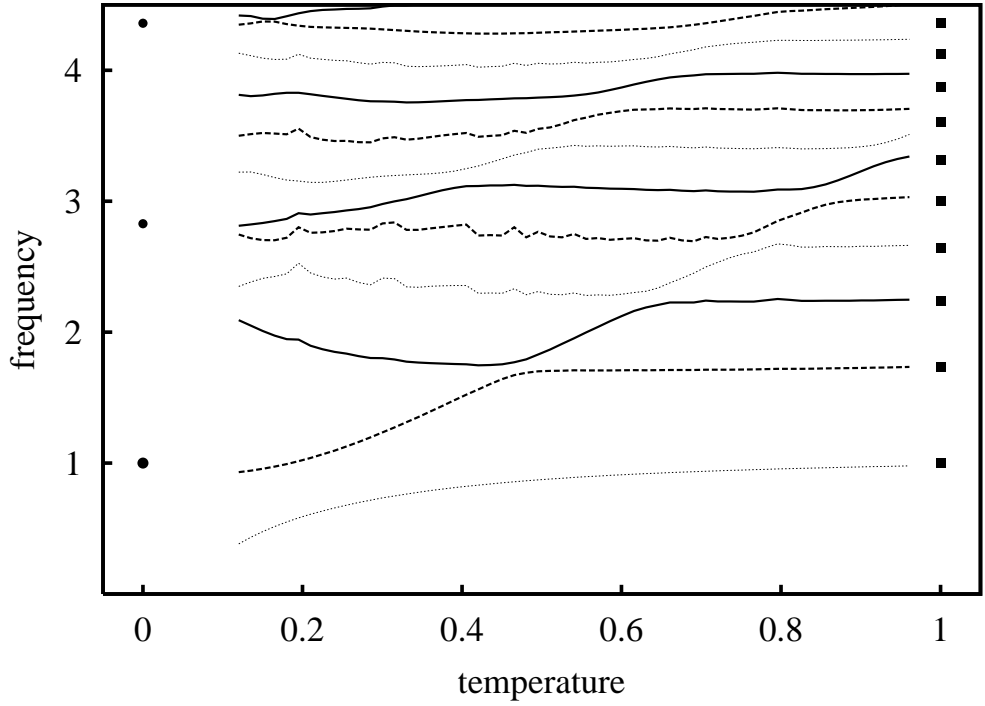}
\caption{Eigenfrequencies versus temperature for $l=1$.}
\label{rys42}
\end{figure}

Figures \ref{rys41} and \ref{rys42} are the main results of this paper.
They show temperature dependence of the low-energy excitation spectrum
for different values of orbital quantum number. For $T=0$ and $T=T_c$
frequencies are known analytically: dots represent the spectrum
\cite{Stringari}:
\beq
\omega = \sqrt{2 n^2 + 2n l + 3n + l}
\eeq
and squares the spectrum (\ref{3widchm}).
It can be seen that numerical results are in fair agreement with these
spectra. The "horizontal" lines are the frequencies where
thermal cloud oscillation is non-zero at all temperatures;
for $T=0$ where there is no thermal cloud, those frequencies are
not valid anymore. A complementary set of lines (inclined) originates
from the $T=0$ condensate frequencies.

\begin{figure} \centering
\begin{tabular}{@{}cc@{}}
\epsfxsize=7.8cm
\epsffile{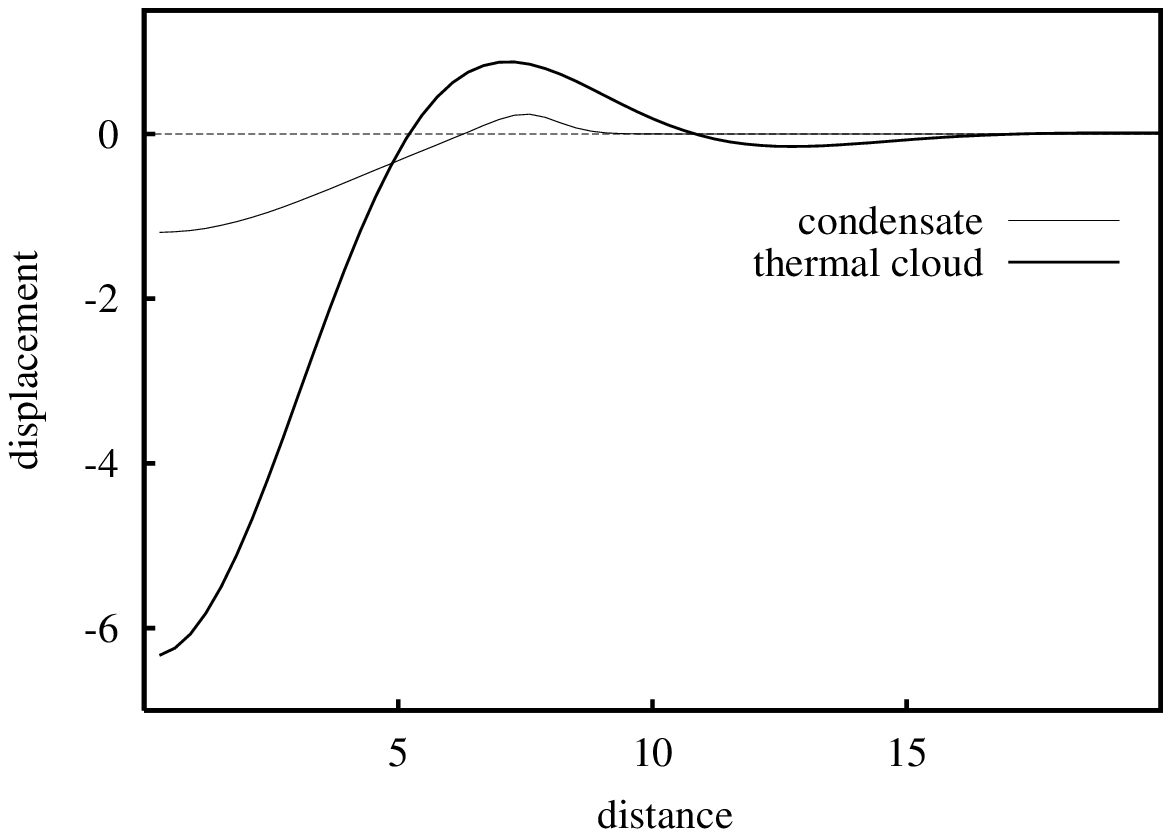} &
\epsfxsize=7.8cm
\epsffile{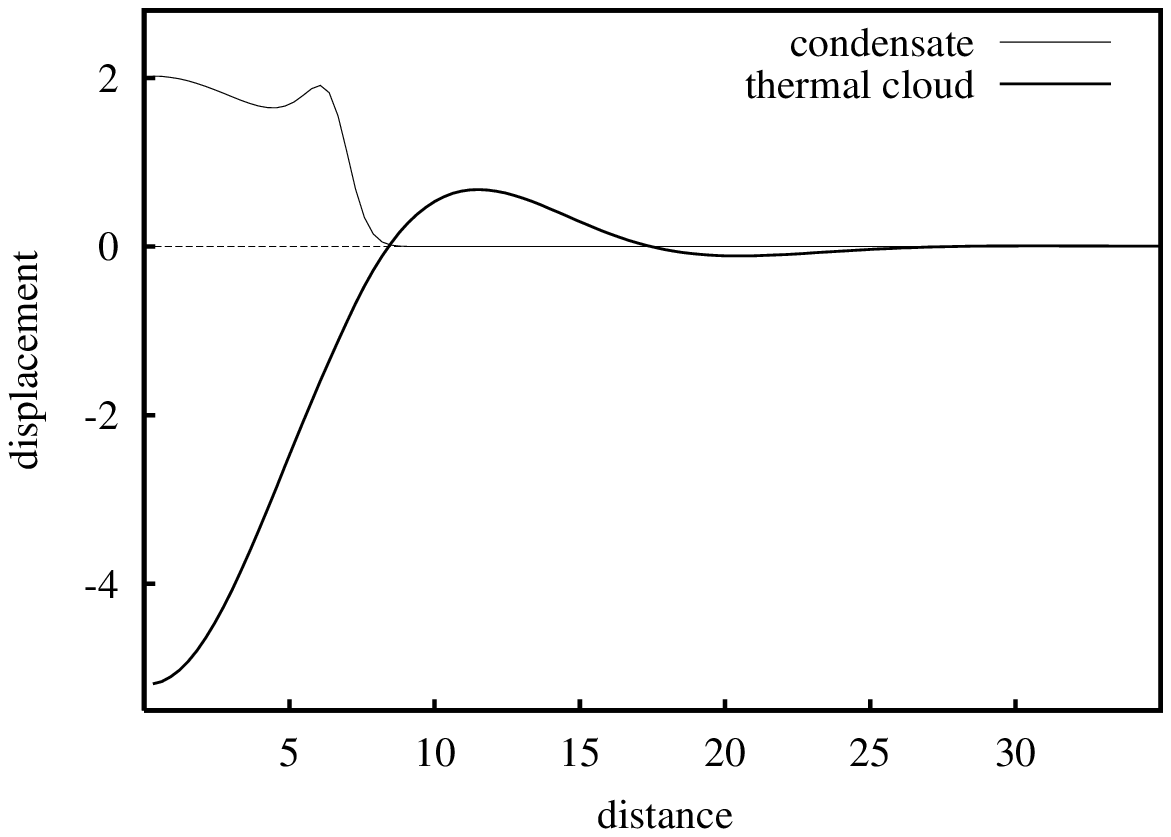} \\
\end{tabular}
\caption{Shapes of eigenmodes.}
\label{rys43}
\end{figure}

For the temperatures between $T=0$ and $T=T_c$ both phases oscillate
with similar amplitudes. Examples are plotted in figure \ref{rys43}.

The existence of the avoided crossings in the spectrum in figures \ref{rys41}
and \ref{rys42} is clear. They occur when two eigenmodes as a function of
a continuous parameter approach each other, and after a certain point they
start to diverge.

\begin{figure} \centering
\begin{tabular}{@{}cc@{}}
\epsfxsize=7.8cm
\epsffile{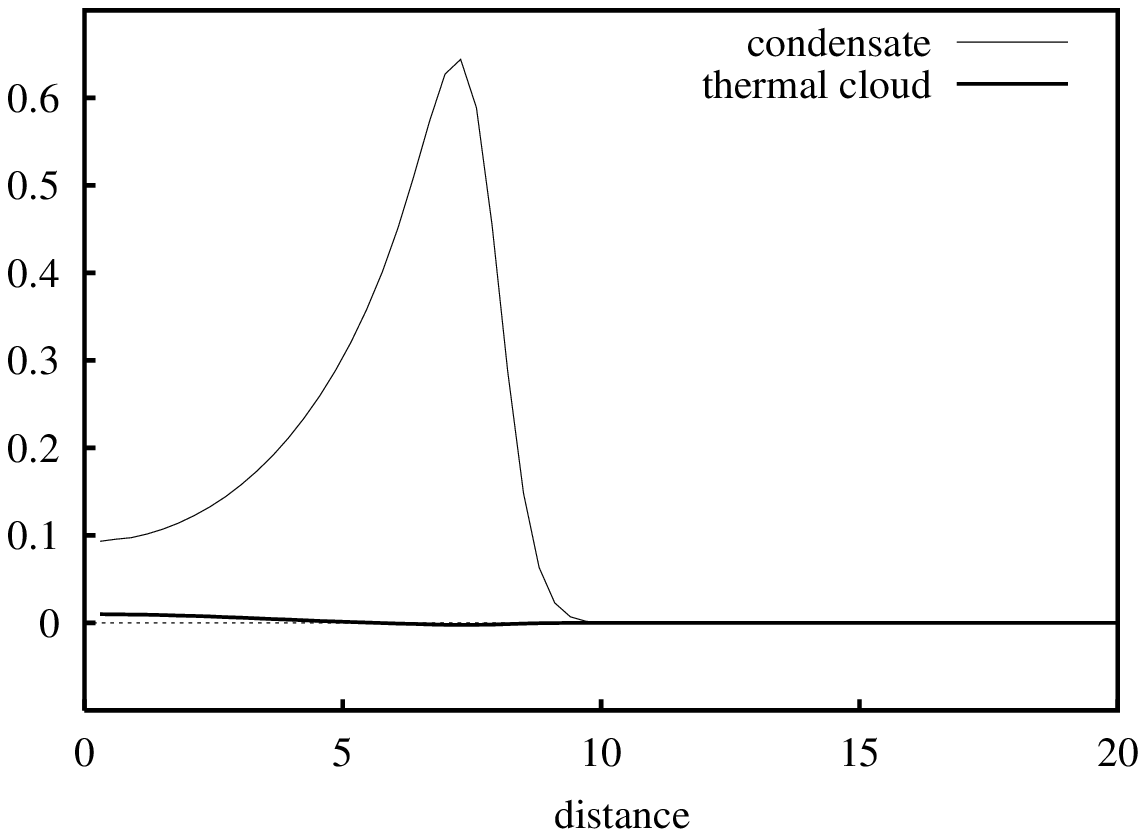} &
\epsfxsize=7.8cm
\epsffile{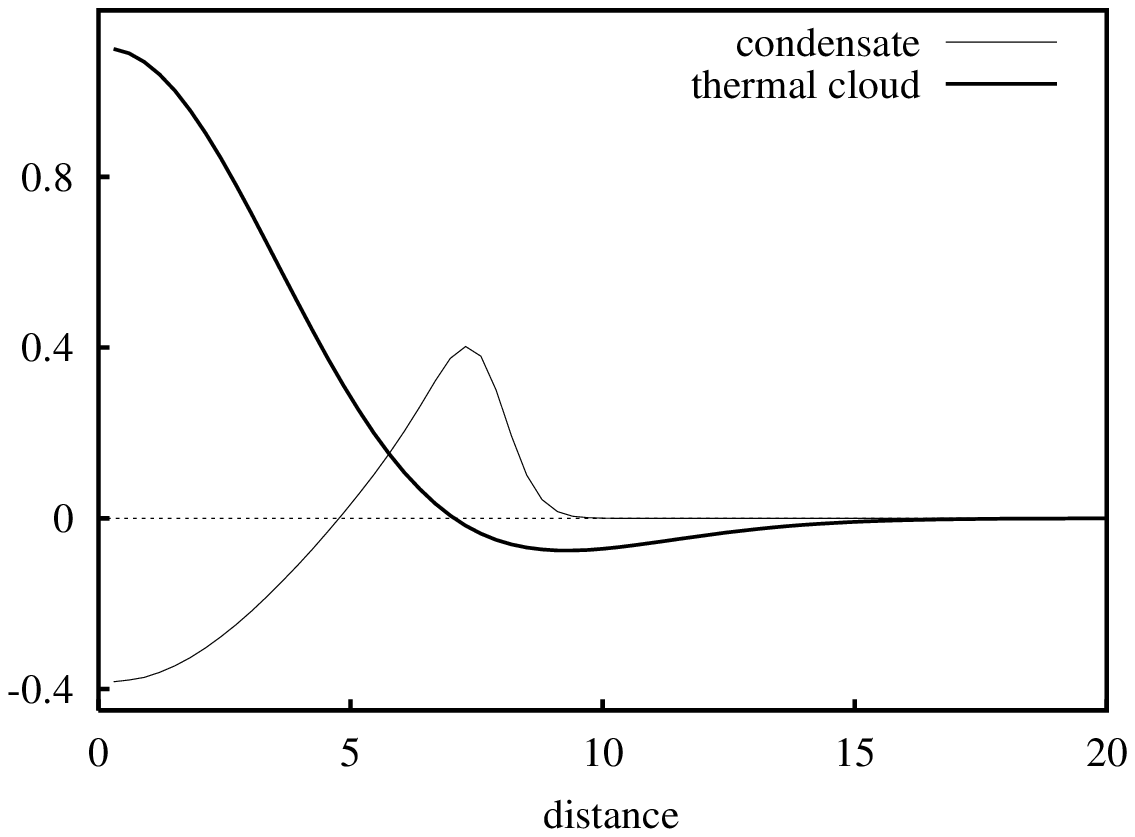} \\
\epsfxsize=7.8cm
\epsffile{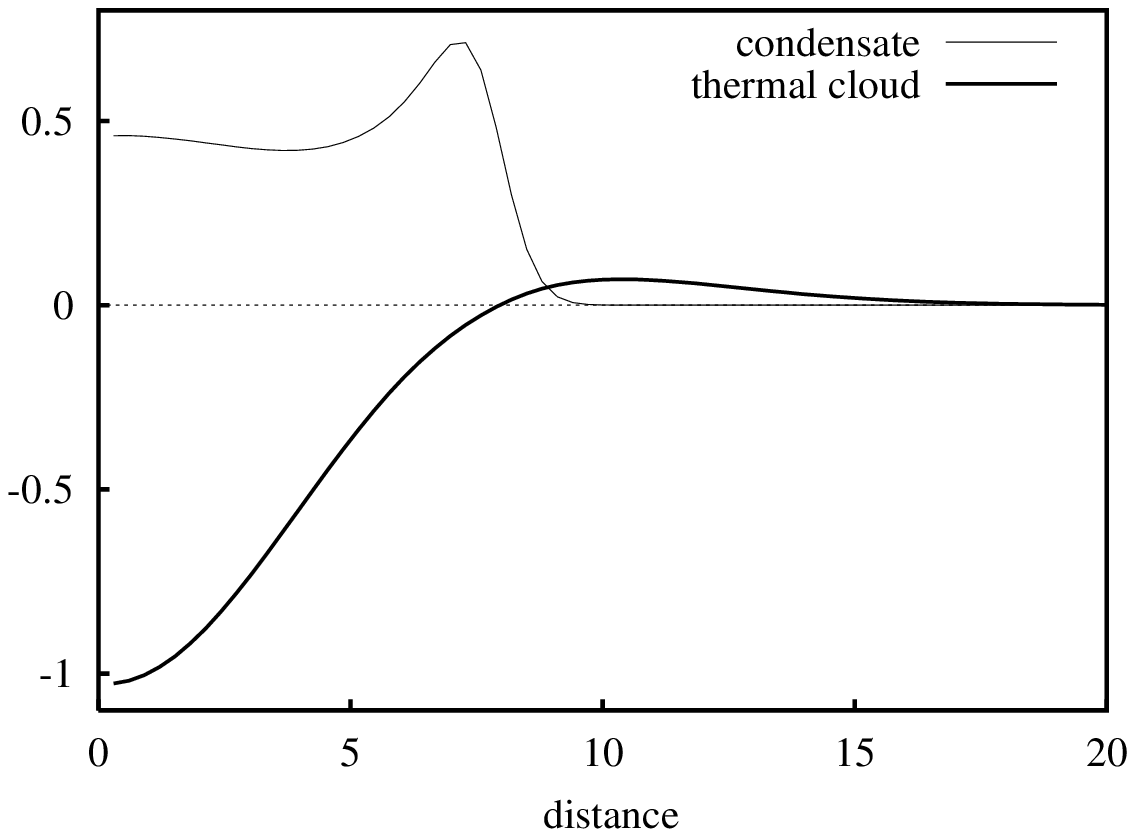} &
\epsfxsize=7.8cm
\epsffile{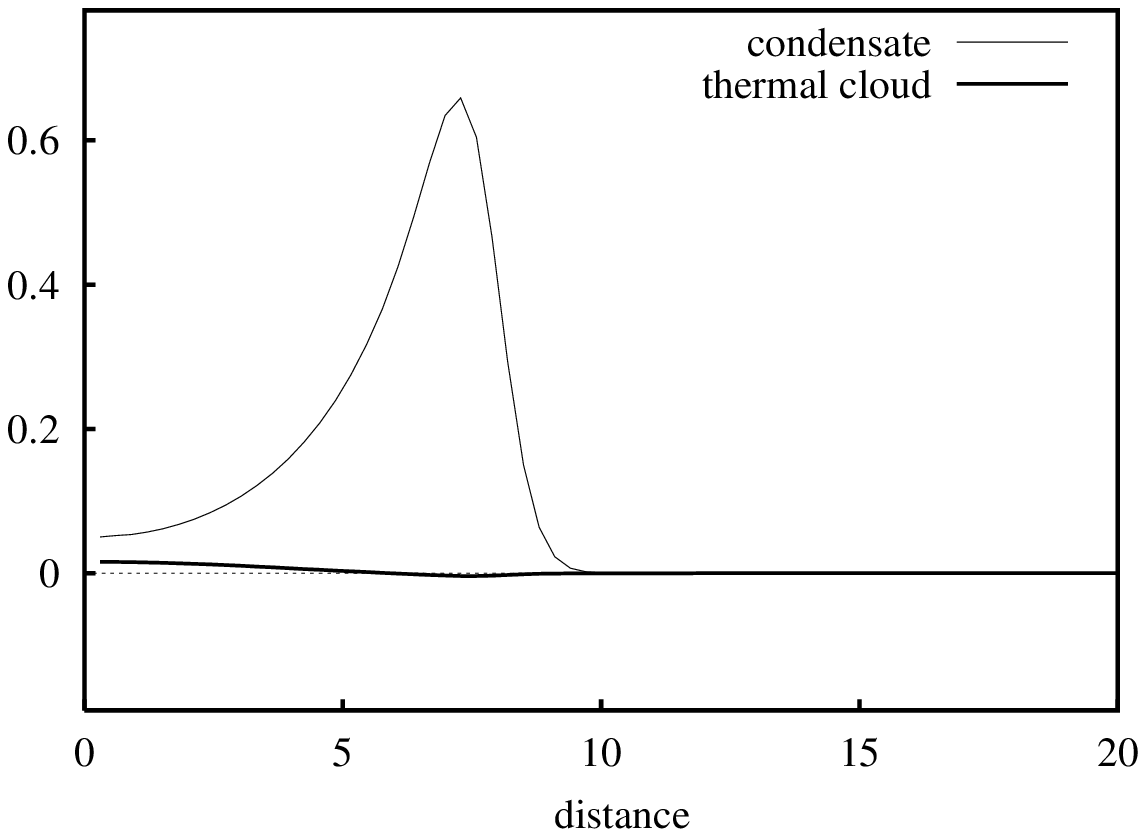} \\
\end{tabular}
\caption{Shapes of the eigenmodes around an avoided crossing.}
\label{rys45}
\end{figure}

Avoided crossing is a well known phenomenon in physics. Its general properties
are described by the Landau-Zener theory. According to it, the character of
eigenfunction "jumps" across the avoided crossing, as seen on the figure
\ref{rys45}. If the parameter is time, two types of processes might be
recognized: adiabatic (slow) and diabatic (quick). The time scale
distinguishing between those types is connected to the minimal distance
between the eigenenergies -- via the uncertainty principle.

It would be interesting to observe these avoided crossings experimentally.
Unfortunately, due to particular realization of experiments, those
processes cannot be observed, because temperature is a parameter given
from outside, not changing during the experiment. However, our conjecture
is that the observed rapid dependencies of the eigenfrequencies with
temperature occur because of the avoided crossings.

It is also clear that due to the interaction with the thermal cloud
mode which in $T=0$ had frequency $\omega=0$ is awakening and no
longer has it zero frequency. Therefore it has a potential of being
experimentally observed.

\section{Summary and Acknowledgments}
We investigated a simplified version of the two-gas model. Within this
model dynamic interactions between the phases: condensate and thermal
cloud, are included. No mechanism for exchanging particles nor
for damping is provided, though.

Ground state of the system is found numerically using a propagation
in imaginary time technique. Equations resulting from small oscillations
theory constitute an eigenproblem, which is numerically solved for
spherically symmetric trap.

The resulting spectrum of frequencies exhibits interesting features.
Firstly, it is composed of lines originating from pure condensates
oscillations in $T=0$ and from thermal cloud oscillations ($T=T_c$).
In wide range of temperatures, however, both phases oscillate
with approximately the same amplitudes. Secondly, avoided crossings
are observed. We suggest that experimentally observed sudden shifts
in frequencies might be a result of an avoided crossing. Finally,
due to the interaction with the thermal cloud the awakening of mode
$n=0$ is observed, and might be experimentally
detected. To make contact with the existing experimental work
we plan a similar study in the axially symmetric geometry.

We would like to thank C.Clark for fruitful discussions. R.B. and
K.R. acknowledge the support of the subsidy from the Foundation
for Polish Science. M.B. and K.R. thank for the support of
MCS Grant PAN/NIST-98-340.


\begin{thebibliography}{}

\bibitem{Cornell:1} D.S. Jin, J.R.Enscher, M.R.Matthews, C.E.Wieman, and E.A.Cornell,
 Phys. Rev. Lett. {\bf 77}, 420 (1996).
\bibitem{Ketterle:1} M.-O. Mewes, M.R. Andrews, N.J. van Druten, D.M. Kurn,
D.S. Durfee, C.G. Townsend, and W. Ketterle, Phys. Rev. Lett. {\bf
77}, 988 (1996).

\bibitem{Clark} M. Edwards, P.A. Ruprecht, K. Burnett, R. Dodd, and C.W. Clark,
Phys. Rev. Lett., {\bf 77}, 1671 (1996).
\bibitem{You} L. You, W. Hoston, and M. Lewenstein, Phys. Rev. A {\bf 55},
1581 (1997).
\bibitem{Stringari} S. Stringari, Phys. Rev. Lett. {\bf 77}, 2360
(1996).
\bibitem{Schlapnikow} P. {\"O}hberg, E.L. Surkov, I. Tittonen, S. Stenholm, M. Wilkens, and
G.V. Schlyapnikov, Phys. Rev. A {\bf 56}, 3346 (1997).

\bibitem{Cornell:2} D.S. Jin, M.R. Matthews, J.R. Ensher, C.E. Wieman, and E.A.Cornell,
Phys. Rev. Lett. {\bf 78}, 764 (1997).
\bibitem{Ketterle:2} D.M. Stamper-Kurn, H.-J. Miesner, S.Inouye, M.R. Andrews, and W. Ketterle,
Phys. Rev. Lett. {\bf 81}, 500 (1998).

\bibitem{Popow} V. N. Popow, {\em Funcional Integrals and Collective Modes}
(Cambridge University Press, New York, 1987).
\bibitem{Hutchison} D.A.W. Hutchinson, E. Zaremba, and A. Griffin,
Phys. Rev. Lett. {\bf 78}, 1842 (1997).
\bibitem {Dodd:1} R.J. Dodd, M. Edwards, C.W. Clark, and
K. Burnett, Phys. Rev. A {\bf 57}, R32 (1998).
\bibitem{Dodd} D.A.W. Hutchinson, R.J. Dodd, and  K. Burnett,
Phys. Rev. Lett. {\bf 81}, 2198 (1998).
\bibitem{Stoof} M.J. Bijlsma and H.T.C. Stoof, Phys. Rev. A {\bf 60}, 3973
(1999).
\bibitem{Griffin:1} E. Zaremba, A. Griffin, and T. Nikuni, Phys. Rev. A {\bf 57}, 4695
(1998).

\bibitem{Rzaz} P. Navez, D. Bitouk, M. Gajda, Z. Idziaszek and K. Rz{\as}{\.z}ewski,
Phys. Rev. Lett. {\bf 79}, 1789 (1997).
\bibitem{fse:1} W. Ketterle and N.J. van Druten, Phys. Rev. A {\bf 54}, 656
(1996).
\bibitem{fse:2} S. Giorgini, L.Pitaevskii, and S.Stringari, Phys. Rev. A
{\bf 54}, 4633 (1996).

\bibitem {2gas:1} R.J. Dodd, K. Burnett, M. Edwards, and
C.W. Clark, J. Phys. B {\bf 32}, 4107 (1999).
\bibitem {2gas:2} R.J. Dodd, K. Burnett, M. Edwards, and
C.W. Clark, Acta Phys. Pol. A {\bf 93}, 45 (1998).
\bibitem{Griffin:2}  A. Griffin, Phys. Rev. B {\bf 53}, 9341
(1996).
\bibitem{Madelung} E. Madelung, Z. Phys. {\bf 40}, 322 (1926).

\bibitem{Abramowicz} M. Abramowitz and I.A. Stegun, {\em Handbook of mathematical
functions} (Dover Publications, New York, 1965).

\end{thebibliography}
\end{document}